\documentclass[epj]{svjour}
\usepackage{graphicx,bbm,amsmath,amssymb,verbatim,times}
\begin{document}
\authorrunning{P. Aniello, C. Lupo, M. Napolitano, and M G.A. Paris}
\titlerunning{Engineering multiphoton states for linear optics computation}
\title{Engineering multiphoton states for linear optics computation}
\author{Paolo Aniello$^1$, Cosmo Lupo$^1$, Mario Napolitano$^1$,
and Matteo G.A. Paris$^2$} \institute{ $^1$ INFN Sezione di Napoli
and Dipartimento di Scienze Fisiche  dell'Universit\`a di
Napoli `Federico II', C.\ U.\ di Monte S.~Angelo, via Cintia, 80126 Napoli,  Italia. \\
$^2$ Dipartimento di Fisica dell'Universit\`a di Milano, Italia.}
\date{\today}
\abstract{ Transformations achievable by linear optical components
allow to generate the whole unitary group only when restricted to
the one-photon subspace of a multimode Fock space. In this paper, we
address the more general problem of encoding quantum information by
multiphoton states, and elaborating it via ancillary extensions,
linear optical passive devices and photodetection. Our scheme stems
in a natural way from the mathematical structures underlying the
physics of linear optical passive devices. In particular, we analyze
an economical procedure for mapping a fiducial 2-photon 2-mode state
into an arbitrary 2-photon 2-mode state using ancillary resources
and linear optical passive $N$-ports assisted by post-selection. We
found that adding a single ancilla mode is enough to generate any
desired target state. The effect of imperfect photodetection in
post-selection is considered and a simple trade-off between success
probability and fidelity is derived. }
 \maketitle
\section{Introduction}
A quantum computer~\cite{N-C}, although still a chimera as a
concrete device, is already a venerable object for physicists,
mathematicians and computer scientists, for the wide range of
completely new perspectives that such a tool should offer for the
development of science as well as for technological applications.

Photon states are stable against decoherence, and are currently
produced and manipulated in modern laboratories. These features
make the possibility of implementing quantum logic gates
particularly attractive. One of the most promising architectures
for implementing a quantum computer by means of optical systems is
based on a scheme proposed by Knill, Laflamme and Milburn
(KLM)~\cite{KLM}. In this scheme, information is encoded by
(tensor products of) single-photon two-mode states of the
quantized e.m.\ field; precisely, the qubit states are identified
with a couple of single-photon states on two optical modes (dual
rail logic) and multi-qubits are obtained by tensor products. The
basic ingredients for the elaboration of information in the KLM
scheme are linear optical passive (LOP) components~\cite{review}
--- essentially, phase shifters and beam splitters --- by which
one is able to realize the single qubit gates; all other
operations can be obtained in a non-deterministic way exploiting,
in addition, ancillary optical modes and photodetection. One can
show that, with the KLM scheme --- hence, using only single photon
sources, LOP devices and photodetectors
--- it is possible to simulate efficiently, i.e.\ by means of a
polynomial amount of resources, an ideal quantum
computer~\cite{review_1,review_2}.

It is worth noting, however, that in the KLM scheme, states that
are not in the dual rail logic (e.g.\ the state $|2000\rangle$)
may be produced during the computation process, even if at the
output they recombine to get back to the dual rail encoding. As it
will be shown in the following, this is a consequence of the fact
that the linear space spanned by all $n$-photon states (on a given
number $N$ of optical modes) is the carrier Hilbert space of an
irreducible unitary representation of $\mathrm{U}(N)$ which is
associated in a natural way with the action of LOP devices. It
seems then quite natural to investigate, in addition to the KLM
dual rail logic, also the possibility of encoding information by
means of $n$-photon $N$-mode states, with $n\ge 1$ and $N\ge 2$.
The case where $n=1$ and $N\ge 2$, with gates implemented only by
LOP components, has been considered by Cerf \emph{et
al.}~\cite{Cerf}. This scheme is easily seen to be not scalable.

In this paper, we will consider the case where information is
encoded by $n$-photon states, with $n>1$, on $N\ge 2$ modes and
logic gates are obtained by LOP components \emph{and}
photodetectors. As anticipated, this scheme stems in a natural way
from the mathematical structures underlying the physics of LOP
devices, structure that has been investigated in two previous
papers~\cite{P-R,future}. We will now address, as a first step,
the following problem: to engineer any desired state --- which may
be regarded as the `input state' of a quantum computation process
--- in the chosen encoding space, starting with a fixed `fiducial
state', namely, a photon state that can be easily produced by
single-photon sources. For the sake of definiteness, we will focus
on the case where $n=N=2$. This is the simplest case that is not
contemplated in the Cerf \emph{et al.}\ and in the KLM schemes.
Notice that in our case the building blocks of quantum information
are \emph{qutrits} instead of qubits.

The paper is structured as follows. In
Section~\ref{group-section}, the basic mathematical ingredients
for a natural and systematic description of LOP transformations
are recalled. Next, in Sections~\ref{twomodes-section}
and~\ref{post-section}, our encoding and elaboration scheme is
presented. The effect of realistic imperfect photodetection is
then discussed in Section~\ref{real-section}. Finally, in
Section~\ref{outro}, we end up with some concluding remarks.
\section{A group theoretical approach to LOP components}
\label{group-section} 
\begin{figure}[h]
\centering
\includegraphics[width=0.2\textwidth]{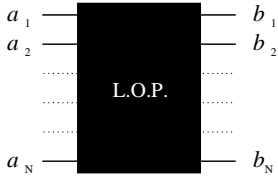}
\caption{a black-box picture of a $2N-$port based on L.O.P. transformation}
\label{2Nport}
\end{figure}

A generic LOP transformation can be described
as a $2N-$port, namely a \emph{black box} with $N$ input modes and
$N$ output modes. A pictorial representation is given in figure
\ref{2Nport}. In two previous papers \cite{P-R,future}, it has been
shown that the natural mathematical description of the action of LOP
transformations is based on the theory of representations of
semi-simple Lie groups and algebras; in this framework a special
role is played by the Jordan-Schwinger map \cite{JSmap,JSmap_1}. In
this section, we recall the basic ingredients of such a description.
Let us consider a set of $N$ optical modes with the associated field
operators
\begin{equation}
a_i, a_i^\dag \ \ \ \ i=1,2\ldots N,
\end{equation}
where the index $i$ may label both spatial or polarization
modes of the field, with the canonical commutation relations
\begin{equation} \label{CCR}
[a_i, a_j^\dag] = \delta_{ij}{\mathbbm I}, \ \ \ \ [a_i, a_j] =
[a_i^\dag, a_j^\dag] = 0,
\end{equation}
\sloppy where ${\mathbbm I}$ is identity operator. It is well known
that the set of operators $\{ a_i, a_i^\dag, {\mathbbm I} \}$,
endowed with the canonical commutation relations (\ref{CCR}), are
the generators of a realization of the $N-$dimensional
Heisenberg-Weyl algebra $\mathcal{W}(N)$ \cite{P-R}. We indicate
with $\mathcal{H}^{(N)}$ the bosonic Fock space associated with the
chosen set of $N$ modes.
\par
We are interested in Linear Optical Passive (LOP) transformations,
 \emph{i.e.}, maps that are linear in the field amplitudes
\fussy
\begin{eqnarray} \label{linear}
\left\{
\begin{array}{ccc}
a_i      & \longrightarrow & b_i = M_{ij} a_j + N_{ij} a_j^\dag           \\
a_i^\dag & \longrightarrow & b_i^\dag = M_{ij}^* a_j^\dag + N_{ij}^* a_j
\end{array}\right.
\end{eqnarray}
(where the sum over repeated indices is assumed) and preserve the
total photon number operator
\begin{equation} \label{number}
\sum_{i=1,\ldots N} b_i^\dag b_i = \sum_{i=1,\ldots N} a_i^\dag a_i.
\end{equation}
It is easy to show that the only maps with
properties (\ref{linear}-\ref{number})
are of the form:
\begin{eqnarray}\label{unitary}
\left\{
\begin{array}{ccc}
a_i      & \longrightarrow & b_i = M_{ij} a_j             \\
a_i^\dag & \longrightarrow & b_i^\dag = M^*_{ij} a_j^\dag
\end{array}\right.
\end{eqnarray}
where $M_{ij}$ is a $N \times N$ unitary matrix ($M \in
\mathrm{U}(N)$). It is also a simple calculation to verify that a
map of the form (\ref{unitary}) preserves the canonical commutation
relations:
\begin{eqnarray}
&[b_i, b_j^\dag] = [a_i, a_j^\dag] = \delta_{ij}{\mathbbm I}, &\\
&[b_i, b_j] = [b_i^\dag, b_j^\dag] = [a_i, a_j] = [a_i^\dag,
a_j^\dag] = 0.&
\end{eqnarray}
\sloppy Thus one can consider the two realizations of the ($N$
dimensional) Heisenberg-Weyl algebra given by $\{ a_i, a_i^\dag,
{\mathbbm I} \}$ and $\{ b_i, b_i^\dag, {\mathbbm I} \}$ and notice
that by virtue of the Stone-von Neumann theorem \cite{Stone,vonN}
they are unitarily equivalent, that is, it exists an unitary
operator $U$ acting in the $N-$modes Fock space such that
\begin{eqnarray}
\label{similarity} \left\{
\begin{array}{ccc}
b_i      &=& U^\dag a_i U          \\
b_i^\dag &=& U^\dag a_i^\dag U
\end{array}\right.
\end{eqnarray}
Notice that the operator $U$ is defined only up to an arbitrary
phase factor. Since by construction $U$ commutes with the total
photon number operator this phase factor is fixed by the action of
$U$ on the vacuum state:
\begin{equation} \label{phase.factor}
U |0\rangle = e^{i\phi(U)} |0\rangle.
\end{equation}
This ambiguity can be removed if one considers an explicit
construction of the unitary operator $U$. This can be done by means
of the Jordan-Schwinger map. \fussy
\subsection{The Jordan-Schwinger map}
The Jordan-Schwinger (JS) map \cite{JSmap,JSmap_1}, in its general
formulation, maps a Lie algebra into an algebra of operators defined
on a bosonic Fock space, this map being an algebra homomorphism. The
JS map is defined as follows. Let us consider a $m-$dimensional Lie
algebra realized as an algebra of $N \times N$ matrices with a given
basis of generators
\begin{equation} \label{geners}
Q^{(\alpha)} \equiv ||Q^{(\alpha)}_{ij}|| \ \ \ \ \alpha =
1,2,\ldots m \ \ i,j=1,2,\ldots N
\end{equation}
and commutation relations
\begin{equation}
[ Q^{(\alpha)}, Q^{(\beta)} ] = c^{\alpha\beta}_{\gamma}
Q^{(\gamma)}.
\end{equation}
Let us also consider a $N-$mode bosonic Fock space with field
operators $a_i, a_i^\dag \ \ i=1,2,\ldots N$ and the (normal
ordered) operators
\begin{equation} \label{dijs}
d_{ij} = a_i^\dag a_j.
\end{equation}
The operators (\ref{dijs}) satisfy the following
commutation relations:
\begin{equation} \label{dijs_CR}
[ d_{ij} , d_{hk} ] = d_{ik} \delta_{hj} - d_{hj} \delta_{ik}.
\end{equation}
One can consider the following set of bosonic operators:
\begin{equation}
JS(Q^{(\alpha)}) = Q^{(\alpha)}_{ij} d_{ij}.
\end{equation}
The map defined on the basis (\ref{geners})
\begin{equation}
Q^{(\alpha)} \longrightarrow JS(Q^{(\alpha)}) = Q^{(\alpha)}_{ij}
d_{ij}\;,
\end{equation}
extended by linearity, defines the JS map. It is easy to show that
by virtue of the commutation relations (\ref{dijs_CR}) the JS map is
indeed an algebra homomorphism, namely
\begin{equation}
[ JS(Q^{(\alpha)}), JS(Q^{(\beta)}) ] = c^{\alpha\beta}_{\gamma}
JS(Q^{(\gamma)}).
\end{equation}
Let us now come back to the transformation (\ref{similarity}). The
$N \times N$ unitary matrix $M$ can be written in terms of the
exponential map as $M = \mbox{exp}(A)$, where $A$ is an element of
the Lie algebra of the $N-$dimensional unitary group (namely a $N
\times N$ anti-hermitian matrix). It is found that the related
unitary operator $U$ can be written in the following way by
exploiting the JS and the exponential map, namely
\begin{equation}
U = \mbox{exp}(JS(A)).
\end{equation}
In order to check this, consider that, for $\epsilon << 1$, we
have:\footnote{For a rigorous proof one can use the well known
formula $e^X Y e^{-X} = \mbox{exp}(\mathrm{ad}_X) Y$, for linear
operators $X$ and $Y$.}
\begin{align}
U^\dag a_k U & = \mbox{exp}(-\epsilon JS(A)) a_k \mbox{exp}(\epsilon JS(A)) \nonumber \\
             & \sim a_k + \epsilon [a_k,JS(A)]
\end{align}
where
\begin{align}
[a_k,JS(A)] & = A_{ij} \left( a_k a_i^\dag a_j - a_i^\dag a_j a_k
\right) \nonumber
\\ &=
A_{ij} \left( (a_i^\dag a_k + \delta_{ik}) a_j - a_i^\dag a_j a_k
\right) = A_{kj} a_j.
\end{align}
The JS map allows to fix the arbitrary phase factor in
(\ref{phase.factor}). In fact, since the $JS(A)$ is a normally
ordered operator, we have:
\begin{equation}
\mbox{exp}(JS(A))|0\rangle = |0\rangle,
\end{equation}
so that $e^{i\phi(U)}=1$.

To summarize we have shown that LOP transformations on $N$ modes are
described by means of the $N-$di\-men\-sio\-nal unitary group acting
on field operators as in (\ref{unitary}). Such an action of the
$N-$di\-men\-sio\-nal unitary group induces a bosonic representation
of the group $\mathrm{U}(N)$ acting on the $N$ modes bosonic Fock
space
\begin{equation}
U = \Upsilon^{(N)}(M)
\end{equation}
that can be explicitly defined by means of the JS map. Indeed it is
easy to check that
\begin{equation}\label{anti-rep}
\Upsilon^{(N)}(M_1 M_2) = \Upsilon^{(N)}(M_1) \Upsilon^{(N)}(M_2).
\end{equation}

Since, by construction, $\Upsilon^{(N)}(M)$ commutes with the total
photon number operator, the unitary representation $\Upsilon^{(N)}$
can be written as the direct sum of unitary (sub) representations
acting on the subspaces with fixed photon number
\begin{equation}
\Upsilon^{(N)} = \bigoplus_{n=0,1,\ldots \infty} \Upsilon^{(N)}_n,
\end{equation}
in correspondence with the decomposition
\begin{equation}
\mathcal{H}^{(N)} = \bigoplus_{n=0,1,\ldots \infty}
\mathcal{H}^{(N)}_n,
\end{equation}
where $\mathcal{H}^{(N)}_n$ is the subspace with $n$ photons on $N$
optical modes. A simple calculation shows that the following
relation holds:
\begin{equation}
\mbox{dim} \mathcal{H}^{(N)}_n = \frac{(n+N-1)!}{n!(N-1)!}.
\end{equation}
Hence, the subspace $\mathcal{H}^{(N)}_n$ can be seen as the space
of a qu-$d$it with $d = \mbox{dim} \mathcal{H}^{(N)}_n$. The (sub)
representation with $n=0$ is the trivial representation of
$\mathrm{U}(N)$:
\begin{equation}
\Upsilon^{(N)}_0(M) |0\rangle = |0\rangle;
\end{equation}
while a special role is played by the $n=1$ (sub) representation
since
\begin{align}
\langle 0| a_k U a_l^\dag |0\rangle & =\langle 0|U (U^\dag a_k U)
a_l^\dag |0\rangle = \langle 0|b_k a_l^\dag |0\rangle \nonumber
\\ & = \sum_m
M_{kh}\langle 0|a_h a_l^\dag |0\rangle = M_{kl},
\end{align}
where we have used the fact that $U U^\dag={\mathbbm I}$ and
$U^\dag|0\rangle =U|0\rangle = |0\rangle$. The nature of the
representations $\{ \Upsilon^{(N)}_n \}_{n=0,1,\ldots}$ has been
studied in detail in \cite{P-R,future}. A remarkable result is that
each (sub) representation $\Upsilon^{(N)}_n$ is a irreducible
unitary representation (IUR) of the group $\mathrm{U}(N)$. For $N=2$
and $n=1$, $\Upsilon^{(2)}_1$ is the relevant (sub) representation
for the implementation of single qubit gates in the framework of
dual rail logic \cite{KLM,P-R}.
\section{About two-mode multiphoton states} \label{twomodes-section}
For the sake of definiteness, in the following we will focus on the
case where $N=2$. This configuration is at the basis of the KLM
scheme for a quantum computer \cite{KLM}, in which the qubit Hilbert
space is identified with the space $\mathcal{H}^{(2)}_1$ of one
photon on two modes. An important requirement for a well defined
quantum computation is the ability to perform an arbitrary one qubit
gate \cite{DiVi}, that is, a generic unitary transformation in the
qubit space $\mathcal{H}^{(N)}_1$. It is a well known result that in
KLM scheme every one qubit gates can be implemented with only
two-modes LOP transformations. This follows directly from the fact
that $\Upsilon^{(2)}_1$ is the fundamental representation of the
group $\mathrm{U}(2)$ acting on the one photon subspace
$\mathcal{H}^{(2)}_1$ (with $\mbox{dim} \mathcal{H}^{(2)}_1 = 2$).
This is no more true in those subspaces characterized by a larger
number of photons. For $n \geq 1$, $\Upsilon^{(2)}_n$ is a
spin$-\frac{n}{2}$ representation acting in the $n$ photon subspace
$\mathcal{H}^{(2)}_n$ \cite{P-R} (with $\mbox{dim}
\mathcal{H}^{(2)}_n = n+1$). Thus, in the case where $n \geq 2$ it
is no more possible to realize a generic unitary gate in the
$n-$photon subspace and, in general, it could not exist a LOP
transformation (associated with some unitary matrix $M$) such that
\begin{equation}
|\psi\rangle = \Upsilon^{(2)}_n(M) |\psi_0\rangle,
\end{equation}
for a \emph{generic} couple of normalized state vectors
$|\psi_0\rangle, |\psi\rangle \in \mathcal{H}^{(2)}_n$; in other
words, for $n \geq 2$ not all the normalized vectors belong to the
same $\mathrm{U}(2)-$orbit.
\par
Let us recall that, given a representation $\Upsilon$ of a group $G$
in a Hilbert space $\mathcal{H}$, the orbit $\mathcal{O}_{\psi_0}$
of the group passing through a given vector
$|\psi_0\rangle\in\mathcal{H}$ is defined as the set of all vectors
$|\psi\rangle\in\mathcal{H}$ such that $|\psi\rangle = \Upsilon(g)
|\psi_0\rangle$, for some $g \in G$. In the case where $n=1$, the
orbit of the group $\mathrm{U}(2)$ in $\mathcal{H}^{(2)}_1$ passing
through a vector of unit norm fulfills the whole unit sphere in
$\mathcal{H}^{(2)}_1$. In the multiphoton case, the orbit of the
group $\mathrm{U}(2)$, acting in $\mathcal{H}^{(2)}_n$ via the
representation $\Upsilon^{(2)}_n$, passing through a normalized
state vector $|\psi_0\rangle \in \mathcal{H}^{(2)}_n$, is only a
proper sub-manifold of the unit sphere.
\par
\sloppy In order to illustrate these arguments explicitly, let us
consider a generic $\mathrm{SU}(2)$ matrix:
\begin{eqnarray} \label{underU}
M = \left[\begin{array}{cc}
\alpha   & \beta    \\
-\beta^* & \alpha^*
\end{array}\right],
\end{eqnarray}
with $\alpha=e^{i\chi} \cos{\theta}$, $\beta=e^{i\phi}
\sin{\theta}$. The two-photon subspace $\mathcal{H}^{(2)}_2$ has
dimension $d=3$, hence it can be seen as a qutrit space. In the
number basis $\{ |20\rangle, |11\rangle, |02\rangle \}$ the operator
$\Upsilon^{(2)}_2(M)$ has a matrix representation
\begin{eqnarray}
\Upsilon^{(2)}_2(M) \equiv
\left[\begin{array}{ccc}
\alpha^2               & \sqrt{2}\alpha\beta      & \beta^2         \\
-\sqrt{2}\alpha\beta^* & |\alpha|^2-|\beta|^2     & \sqrt{2}\alpha^*\beta  \\
{\beta^*}^2            & -\sqrt{2}\alpha^*\beta^* & {\alpha^*}^2
\end{array}\right].
\end{eqnarray}
It explicitly shows that it is not possible to realize every
(qutrit) unitary transformations. Also notice that
\begin{equation}
\Upsilon^{(2)}_2(M)|11\rangle = \sqrt{2}\alpha\beta|20\rangle +
\left(|\alpha|^2-|\beta|^2\right)|11\rangle
-\sqrt{2}\alpha^*\beta^*|02\rangle,
\end{equation}
from which it is apparent that the state vectors $|11\rangle$ and
$|20\rangle$ (or $|02\rangle$) do not belong to the same orbit.
\fussy
\par
To summarize, in the multiphoton case two categories of problems
arise that are not present in the single photon case: 1) given a
state vector $|\psi_0\rangle$, there is in general no LOP
transformation that allows to map $|\psi_0\rangle$ into an arbitrary
\emph{target} state vector $|\psi\rangle$; 2) it is not possible to
perform every qutrit unitary gate only with two-mode LOP
transformations. The latter problem was investigated from different
points of view in \cite{powers,eff-unit,knill_2}. These problems are
related to the DiVincenzo's criteria \cite{DiVi} for a well defined
quantum computation, namely the point 1) is related to the
\emph{ability to initialize the state of the} qutrit \emph{to a
simple fiducial state}; the point 2) is related to the ability to
perform a \emph{universal set of quantum gates}. In the following
sections, we consider the first problem and suggest a solution based
on photodetection on ancillary modes and conditional post-selection.
\section{Projection via a post-selection protocol} \label{post-section}
A remarkable property of IURs is that every orbit is total
\cite{total}. This means that given a normalized \emph{target} state
vector $|\phi\rangle\in\mathcal{H}^{(N)}_n$ and an orbit
$\mathcal{O}_{\psi_0}$ of the IUR $\Upsilon^{(N)}_n$, it is always
possible to find a $|\psi\rangle\in\mathcal{O}_{\psi_0}$ with a non
vanishing projection along $|\phi\rangle$:
\begin{equation}
\langle\phi | \psi\rangle \neq 0.
\end{equation}
While the existence of a non-vanishing projection follows from the
properties of the irreducible representations $\Upsilon^{(N)}_n$,
how to realize physically (at least in principle) such a projection
is a matter of a different nature. In the following we discuss with
some examples a procedure based on photodetection on ancillary
optical modes and conditional post-selection. The result is a
non-deterministic protocol that allows to map a fixed \emph{input}
state into a desired \emph{target} state with a certain probability.
In order to illustrate the idea, let us consider the case of a
qutrit encoded in the subspace with two photons on two modes
$\mathcal{H}^{(2)}_2$. The proposed procedure consists in four
steps. The first step is to initialize the qutrit system in a fixed
input state $|\psi_0\rangle$. The second step is to add one extra
optical mode that plays the role of an ancilla: the state of the
ancillary mode is initialized in a number state with $m$ photons and
we consider the \emph{extended} state
\begin{equation}
|\psi_0\rangle \longrightarrow |\psi_0\rangle|m\rangle.
\end{equation}
Hence, the relevant space for the system+ancilla is
$\mathcal{H}^{(3)}_{2+m}$. The third step is to perform a three-mode
LOP transformation that acts on $\mathcal{H}^{(3)}_{2+m}$ via the
IUR $\Upsilon^{(3)}_{2+m}$.
\begin{equation}\label{decomp}
|\psi_0\rangle|m\rangle \longrightarrow U |\psi_0\rangle|m\rangle =
|\phi_0\rangle|0\rangle + |\phi_1\rangle|1\rangle + \ldots +
|\phi_m\rangle|m\rangle.
\end{equation}
The final step is a post-selection on the ancillary mode: the
\emph{target} state is obtained, with a certain probability
$\mathcal{P}_m$, in correspondence of the detection of $m$ photons
on the ancillary mode.

Overall the transformation of the initial state is described by a
completely positive map ${\cal E}^{(m)}$ which depends on the
initial preparation of the ancilla mode. The Kraus-Sudarshan form of
$\longrightarrow {\cal E}^{(m)}$ is of course given by
\begin{equation}
\label{CPmaps}
|\psi_0\rangle
\langle \psi_0 |
\longrightarrow {\cal E}^{(m)} (|\psi_0\rangle
\langle \psi_0|)=
\sum_{m'}  A_{m'}^{(m)}|\psi_0\rangle\langle\psi_0| A_{m'}^{(m)\dag}\:,
\end{equation}
where $A_{m'}^{(m)}=\langle m' | U | m\rangle$.
The post-selection conditioned on the photodetection of $m'$ photons
on the ancillary mode corresponds to a single branch of the map
{\em i.e.} to the transformation
\begin{equation}
|\psi_0\rangle \longrightarrow A_{m'}^{(m)} |\psi_0\rangle =
|\phi_{m'}\rangle.
\end{equation}
In the following two examples are presented with $m=0,1$. In the
appendix \ref{appx} it was shown that adding one ancillary mode is
indeed sufficient in order to obtain the optimal working point.
\subsection{On the ability to initialize the state of a
qutrit to a simple fiducial state} Let us consider two
\emph{computational} modes with one extra ancillary mode and a
three-mode LOP transformation:
\begin{equation} \label{Matrix}
a_i \longrightarrow b_i = M_{ij} a_j \ \ \ \ i, j = 1, 2, 3.
\end{equation}
Let us also take the third (ancillary) mode initialized in the
vacuum state. Following the procedure outlined above, here we answer
the question of whether is possible to find a three modes LOP
transformation such that, after a photodetection on the third
(ancillary) mode, a generic qutrit state
\begin{equation} \label{target}
|\phi\rangle = A |20\rangle + B |11\rangle + C |02\rangle
\end{equation}
is obtained, with a certain probability, on the first and second
(computational) modes.
\begin{figure}[h]
\centering
\includegraphics[width=0.4\textwidth]{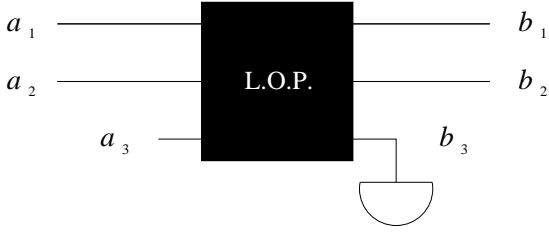}
\caption{Photodetection assisted three-mode L.O.P. transformation}
\label{6port}
\end{figure}
As input state we select the state
$|\psi_0\rangle_{12}=|11\rangle_{12}$ that will be extended with one
ancillary mode initialized in the vacuum state ($m=0$)
\begin{equation}\label{input}
|\psi_0\rangle \longrightarrow |11\rangle_{12}|0\rangle_{3}.
\end{equation}
The subscripts indicate the mode labels and will be omitted in what
follows.

The action of a LOP transformation acting on (\ref{input}) yields
to:
\begin{equation}
|\psi\rangle = M_{p1} M_{q2} a^\dag_p a^\dag_q | 0 0 0 \rangle \ \ \
\ p, q = 1, 2, 3. \label{psiout}
\end{equation}
The global (three-mode) output state, obtained after the three-mode
LOP transformation has the form:
\begin{equation}
|\psi\rangle = |\phi_0\rangle|0\rangle + |\phi_1\rangle|1\rangle +
|\phi_2\rangle|2\rangle,
\end{equation}
where $|\phi_n\rangle$ are two-mode states. A post-selection
conditioned to the vacuum on the third optical mode gives:
\begin{equation}
|\psi\rangle \longrightarrow |\phi_0\rangle |0\rangle = M_{p1}
M_{q2} a^\dag_p a^\dag_q |0 0 \rangle |0\rangle \ \ \ \ p, q = 1, 2
\ \ ,
\end{equation}
where
\begin{eqnarray}
|\phi_0\rangle &=& \sqrt{2} M_{11} M_{12} |20\rangle \nonumber \\
               &+& \left( M_{11} M_{22} + M_{21} M_{12}
               \right)|11\rangle \nonumber \\
               &+& \sqrt{2} M_{21} M_{22} |02\rangle
\label{psi0}
\end{eqnarray}
is the \emph{un-normalized} two-mode output. The square modulus
$\mathcal{P}_0 = \langle \phi_0 | \phi_0 \rangle$ gives the
probability of success of the \emph{vacuum} measurement. From a
mathematical point of view, the question is whether is possible to
find, for every \emph{target} state (\ref{target}), an unitary
matrix $M$ such that the output state (\ref{psi0}) and the target
state (\ref{target}) are equal apart of a normalization (and phase)
factor.
\par
The following propositions hold:
\begin{proposition} \label{P1}
for any $\alpha, \beta, \gamma, \delta$ with $|\alpha|^2+|\beta|^2
\leq 1 $, there exists an unitary matrix
\begin{eqnarray} \label{Mtheorem}
M =  \left[\begin{array}{ccc}
\alpha  & \gamma/k & e_3     \\
\beta   & \delta/k & e_4   \\
e_1     & e_2/k    & e_5
\end{array}\right]
\end{eqnarray}
for some $e_1,\ldots e_5$ and real $k \neq 0$.
\end{proposition}
\textbf{Proof}: in order the matrix $M$ to be unitary the following
equations have to be satisfied:
\begin{eqnarray}
|\alpha|^2 + |\beta|^2 + |e_1|^2 &=& 1              \label{uno}     \\
|\gamma|^2 + |\delta|^2 + |e_2|^2 &=& k^2           \label{due}     \\
|e_3|^2 + |e_4|^2 + |e_5|^2 &=& 1                   \label{tre}     \\
\alpha^* \gamma + \beta^* \delta + e_1^* e_2 &=& 0  \label{quattro} \\
\alpha^* e_3 + \beta^* e_4 + e_1^* e_5 &=& 0        \label{cinque}  \\
\gamma^* e_3 + \delta^* e_4 + e_2^* e_5 &=& 0.       \label{sei}
\end{eqnarray}
Let us suppose that $e_1^* e_2 \neq 0$.
Equation (\ref{uno}) and (\ref{quattro}) yield to
\begin{equation}
|e_2| = \frac{|\alpha^* \gamma + \beta^* \delta|}{\sqrt{1 -
|\alpha|^2 - |\beta|^2}},
\end{equation}
that inserted in (\ref{due}) gives
\begin{equation} \label{kappa}
k^2 = |\gamma|^2 + |\delta|^2 + \frac{|\alpha^* \gamma + \beta^*
\delta|^2}{1-|\alpha|^2-|\beta|^2}.
\end{equation}
Once $k$, $e_1$ and $e_2$ are found, the remaining coefficients can
be easy computed by an orthonormalization algorithm. Otherwise, if
$e_1^* e_2 = 0$, there is always the trivial solution
$e_1=e_2=e_3=e_4=0$ and $e_5=1$. $\Box$ Notice that one can always
choose $\alpha, \beta, \gamma$ and $\delta$ such that the following
normalization condition holds:
\begin{equation} \label{norm-cond}
2|\alpha\gamma|^2 + |\alpha\delta + \beta\gamma|^2 +
2|\beta\delta|^2 = 1.
\end{equation}
The previous proposition implies that starting from the state
$|11\rangle |0\rangle$ for any normalized target state
(\ref{target}) there exists a LOP three-mode transformation such
that, after a post-selection measurement corresponding to the vacuum
on the ancillary mode, the following transformation is obtained
\begin{equation} \label{norma}
|11\rangle |0\rangle \longrightarrow |\phi\rangle|0\rangle.
\end{equation}
\sloppy
\begin{proposition}\label{P2}
$\mathcal{P}_0 = k^{-2}$ is the probability of success of the
post-selection measurement
\end{proposition}
\textbf{Proof}:
within the scheme of figure \ref{6port} the output state is
\begin{equation}
|\phi_0\rangle = \frac{1}{k} \left( \sqrt{2} \alpha \gamma | 2 0
\rangle
               + \left( \alpha \delta + \beta \gamma \right) | 1 1 \rangle
               + \sqrt{2} \beta \delta | 0 2 \rangle \right)
\end{equation}
With the normalization condition (\ref{norm-cond}) we obtain that
$\mathcal{P}_0 = \langle\phi_0|\phi_0\rangle = k^{-2}$ is the
probability of success. $\Box$ Given a normalized state vector in
the form (\ref{target}) the optimal gate corresponds to the maximum
of $\mathcal{P}_0$ (or the minimum of $k^2$) with constraints:
\begin{eqnarray}
\left\{\begin{array}{ccc}
\sqrt{2}\alpha\gamma     & = & A \\
\alpha\delta+\beta\gamma & = & B \\
\sqrt{2}\beta\delta      & = & C
\end{array}\right.
\end{eqnarray}
\fussy
\subsubsection{Examples} \label{examples-section}
\begin{figure}[h]
\centering
\includegraphics[width=0.48\textwidth]{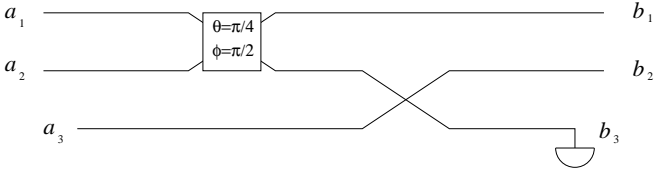}
\caption{Circuital implementation scheme of the three-mode L.O.P.
transformation (\ref{good}) with post-selection procedure}
\label{circuito_1}
\end{figure}

Let us suppose that we want to reach the state $|20\rangle$ starting
from $|11\rangle$:
\begin{equation} \label{imp}
|11\rangle \longrightarrow |20\rangle.
\end{equation}
In the following we are going to describe in which way the
transformation (\ref{imp}) can be obtained with the maximum
probability.  Notice that the state $|20\rangle$ is obtained from
(\ref{target}) taking $B=C=0$, thus we are now looking for
$\mathrm{U}(3)$ matrices of the form
\begin{eqnarray}
M = \left[
\begin{array}{ccc}
\alpha & \gamma/k & e_3 \\
0      & 0        & e_4 \\
e_1    & e_2/k    & e_5
\end{array}\right].
\end{eqnarray}
The normalization condition (\ref{norm-cond}) implies that
$ 2 |\alpha\gamma|^2 = 1 $, and (\ref{kappa}) reads as follows
\begin{equation}
k^{2} = \frac{1}{2} \left( \frac{1}{|\alpha|^2} +
\frac{1}{1-|\alpha|^2} \right).
\end{equation}
Hence, the maximum of probability is $\mathcal{P}_{max} = 1/2$ (that
corresponds to the minimum of $k^2$) and it is reached for
$|\alpha|^2 = 1/2 $. Notice that this is the maximal probability
allowed in the given set up \cite{knill_1}. The corresponding
unitary matrix can be chosen as follows:
\begin{eqnarray} \label{good}
M = \left[\begin{array}{ccc}
\frac{1}{\sqrt{2}} & \frac{i}{\sqrt{2}} & 0 \\
0                  & 0                  & 1 \\
\frac{i}{\sqrt{2}} & \frac{1}{\sqrt{2}} & 0
\end{array}\right].
\end{eqnarray}
This is not the only solution, with this choice the three-mode gate
(\ref{good}) can be decomposed as product of two-mode gates in the
following way
\begin{eqnarray} \label{order}
M = \left[\begin{array}{ccc}
1 & 0 & 0 \\
0 & 0 & 1 \\
0 & 1 & 0
\end{array}\right]
\left[\begin{array}{ccc}
\frac{1}{\sqrt{2}} & \frac{i}{\sqrt{2}} & 0 \\
\frac{i}{\sqrt{2}} & \frac{1}{\sqrt{2}} & 0 \\
0                  & 0                  & 1
\end{array}\right].
\end{eqnarray}
The circuital implementation is schematically represented in figure
\ref{circuito_1} and consists of a symmetric $50\%$ beam splitter on
the first and second mode ($\theta=\pi/4$, $\phi=\pi/2$), followed
by a swap operation between the second and third mode.

As an other example we are going to describe a
post$-$\-se\-lec\-tion assisted LOP transformation with one
ancillary mode which is initialized with one photon. The
computational space is $\mathcal{H}^{(2)}_2$ with an ancillary space
$\mathcal{H}^{(1)}_1$, hence the global space is
$\mathcal{H}^{(3)}_3$. With a procedure analogous to that presented
above, it is easy to shown that the same circuit of equation
(\ref{good}) (and figure \ref{circuito_1}) allows to perform the
transformation
\begin{equation} \label{20_11}
|20\rangle|1\rangle\longrightarrow|11\rangle|1\rangle,
\end{equation}
with an optimal probability of $1/2$.
\section{Effects of imperfect photodetection} \label{real-section}
In the previous sections we have made the assumption that all the
components are ideal: in this section we discuss the presence of
real photodetectors.
Light is revealed by exploiting its interaction with
atoms/molecules or electrons in a solid: each photon ionizes a
single atom or promotes an electron to a conduction band, and the
resulting charge is then amplified to produce a measurable pulse.
In practice, however, available photodetectors are not ideally
counting all photons, and their performances are limited by a
non-unit quantum efficiency $\eta$, namely only a fraction
$\zeta$ of the incoming photons lead to an electric signal, and
ultimately to a {\em count}.  For intense beam of light the
resulting current is anyway proportional to the incoming photon
flux and thus we have a linear detector. On the other hand,
detectors operating at very low intensities resort to avalanche
process in order to transform a single ionization event into a
recordable pulse. This implies that one cannot discriminate
between a single photon or many photons as the outcomes from such
detectors are either a {\em click}, corresponding to any number
of photons, or {\em nothing} which means that no photons have
been revealed.  These {\em Geiger}-like detectors are often
referred to as on/off detectors. For unit quantum efficiency, the
action of an on/off detector is described by the two-value POVM
$\{\Pi_0\doteq |0\rangle\langle 0|, \Pi_1 \doteq {\mathbbm I} -
\Pi_0\}$, which represents a partition of the Hilbert space of
the signal.  In the realistic case, when an incoming photon is
not detected with unit probability, the POVM is given by
\cite{bplis}
\begin{align}
\Pi_0 (\eta) & = \sum_{k=0}^\infty(1-\eta)^k \:
|k\rangle\langle k|, \nonumber \\
\Pi_1 (\eta) & =  {\mathbbm I} - \Pi_0(\eta),\label{onoffPOM}
\end{align}
with $\eta$ denoting
quantum efficiency. As a consequence the conditional state,
occurring when the event "no click" is registered is no longer the
pure state given in Eq. (\ref{psi0}). The conditional state is now
given by the mixed state
\begin{eqnarray}
\varrho_0 &=& \frac{1}{P_0} \hbox{Tr}_3 \left[ |\psi\rangle\langle
\psi| \: {\mathbbm I} \otimes {\mathbbm I} \otimes \Pi_0 (\eta)
\right]
\nonumber \\
&=& \frac{1}{P_0} \sum_{k=0}^2 (1-\eta)^k |\phi_k\rangle\langle
\phi_k | \label{mix0}\;,
\end{eqnarray}
where $|\psi\rangle$ is given in Eq. (\ref{psiout}),
$|\phi_k\rangle$ are the unnormalized states corresponding to an
ideal (unit quantum efficiency, perfect discrimination)
photodetection of $k$ photons and $P_0$ is the global probability of
the "no click" event, {\em i.e}
\begin{eqnarray}
P_0 = \sum_{k=0}^2 (1-\eta)^k \langle\phi_k | \phi_k\rangle.
\label{mixp0}\;
\end{eqnarray}
The (unnormalized) conditional state $|\phi_0\rangle$ is given in Eq.
(\ref{psi0})
whereas $|\phi_k\rangle$, $k=1,2$ are given by
\begin{eqnarray}
|\phi_1\rangle &=& \left( M_{11} M_{32} + M_{31} M_{12}
                   \right) |10\rangle \nonumber \\
               &+& \left( M_{31} M_{22} + M_{21} M_{32} \right) |01\rangle, \\
|\phi_2\rangle &=& \sqrt{2} M_{31} M_{32} |00\rangle
\label{cond12}\;.
\end{eqnarray}
Realistic photodetection thus degrades the quality of the
preparation. In order to asses the whole procedure we use  fidelity
to the target state {\em i.e.}
\begin{eqnarray}
F &=& \frac{1}{\langle\phi_0| \phi_0\rangle} \langle \phi_0 |
\varrho_0 |
\phi_0\rangle \nonumber \\
 &=& \frac{1}{\langle\phi_0| \phi_0\rangle\:P_0} \sum_k  |\langle \phi_0|
 \phi_k\rangle|^2 (1-\eta)^k.
\label{fid}
\end{eqnarray}
Since the conditional states $|\phi_k\rangle$ are mutually
orthogonal we obtain
\begin{equation}
F = \frac{\langle\phi_0|\phi_0\rangle}{\sum_{k=0}^2 (1-\eta)^k
\langle\phi_k | \phi_k\rangle} = \frac{\mathcal{P}_0}{P_0}.
\end{equation}
Therefore there is a simple trade-off between the probability
of success and the quality of the preparation, which can be used
to suitably adapt the procedure to the desired task.
\par
In the case of postselection corresponding to a \emph{click} of
the photodetector the roles of $\Pi_0$ and $\Pi_1$ in
(\ref{onoffPOM}) are inverted. A click
on the ancillary mode corresponds to the preparation of
the computational modes in the mixed states
\begin{equation}
\varrho_1 = \frac{1}{P_1} \sum A_k(\eta) | \phi_k \rangle\langle
\phi_k |,
\end{equation}
where
\begin{equation}
A_k (\eta) = 1 - (1-\eta)^k
\end{equation}
and
\begin{equation}
P_1 = \sum_{k=1}^3 A_k(\eta) \langle\phi_k | \phi_k\rangle.
\end{equation}
The corresponding fidelity to the target state $|\phi_1\rangle$ is
\begin{equation}
F =
\frac{\langle\phi_1|\varrho|\phi_1\rangle}{\langle\phi_1|\phi_1\rangle}
= \frac{1}{\langle\phi_1|\phi_1\rangle P_1} \sum A_k(\eta)
|\langle\phi_1|\phi_k\rangle|^2,
\end{equation}
which simplifies to
\begin{equation}
F = \frac{A_1(\eta)\langle\phi_1|\phi_1\rangle}{\sum_{k=1}^3
A_k(\eta)\langle\phi_k|\phi_k\rangle} = \eta
\frac{\mathcal{P}_1}{P_1}.
\end{equation}
Hence, also in this second example a simple trade off between
probability of success and fidelity of real processes is obtained.
\par
In general, the probability of success and fidelity are independent
quantities in the sense that the maximization of the success
probability does not imply the fidelity optimization. For example,
the optical circuit in figure \ref{circuito_1} corresponds to the
maximal probability of success for both the transformations
$|11\rangle|0\rangle\rightarrow|20\rangle|0\rangle$ and
$|20\rangle|1\rangle\rightarrow|11\rangle|1\rangle$ with an optimal
fidelity for the former and a non-optimal fidelity for the latter.
\section{Conclusive remarks}\label{outro}
In this paper we have addressed the problem of whether in addition
to the KLM dual-rail quantum computation one can consider a more
general $n$-photon $N$-mode encoding scheme; in other words, whether
there is room for quantum information processing based on
multiphoton encoding of {\em qudits}.  In particular, we
investigated the problem of the system initialization in Hilbert
spaces that are carrier spaces of irreducible unitary
representations of unitary groups, representations which are
associated in a natural way with LOP transformations. Focusing on
the case of the 2-photon 2-mode encoding, we found that LOP devices
assisted by post-selection measurements allow to engineer any
desired state in the encoding space starting from a suitable
fiducial state; moreover, we have shown that the use of a single
ancilla mode is enough to ensure the maximum probability of success.
The effects of imperfect photodetection in post-selection have been
considered and a simple trade-off between success probability and
fidelity has been derived.

Of course the lack of further generality and detail in our present
investigation is something to be remedied in the future. However,
we think that it would unrealistic and may be futile, at this
preliminary stage, to try to solve in its full generality the
problem of simulating an ideal quantum computer within the
encoding scheme that we have proposed here. Our main purpose is to
suggest that a deeper understanding of the mathematical structures
underlying LOP devices could be a powerful tool for the further
development of optical quantum computation.

\section*{Acknowledgments}
We wish to thank Prof. G. Marmo of the University of Napoli
`Federico II' for his invaluable scientific and human
support.\\
This work has been supported by MIUR through the project
PRIN-2005024254-002.
\appendix
\section{One ancilla mode is enough} \label{appx}
In the body of the paper we analyzed in some details the preparation
scheme based on a single ancillary mode. In this section we show
that adding a single ancilla is enough in the sense that with
multiple ancillary modes no improvements of the probability of
success can be reached. We consider the case in which the initial
\emph{input} state is the two photon state $|11\rangle$ and discuss
the $m$ ancillary modes generalization of the propositions \ref{P1}
and \ref{P2}. The matrix (\ref{Mtheorem}) has the following
generalized expression in the case of $m$ ancillary modes:
\begin{eqnarray}
M =  \left[\begin{array}{ccc}
\alpha         & \gamma/k         & \mathbf{e_3}     \\
\beta          & \delta/k         & \mathbf{e_4}  \\
\mathbf{e_1}^T & \mathbf{e_2}^T/k & E_5
\end{array}\right],
\end{eqnarray}
where $\mathbf{e_i}$ are $m$-component complex vectors and $E_5$ is
a $m \times m$ matrix.  Equations (\ref{uno}) (\ref{due}) and
(\ref{quattro}) become:
\begin{eqnarray}
\alpha^*\alpha + \beta^*\beta + |\mathbf{e_1}|^2 = 1                              \\
\gamma^*\gamma + \delta^*\delta + |\mathbf{e_2}|^2 = k^2                          \\
\alpha^*\gamma + \beta^*\delta + \langle \mathbf{e_1} , \mathbf{e_2}
\rangle = 0.
\end{eqnarray}
Taking $ \langle \mathbf{e_1} , \mathbf{e_2} \rangle \neq 0$ we obtain
\begin{equation} \label{kappa_2}
k^2 = |\gamma|^2 + |\delta|^2 + \frac{|\alpha^* \gamma + \beta^*
\delta|^2}{|\cos{\theta}|(1-|\alpha|^2-|\beta|^2)},
\end{equation}
where
\begin{equation}
\langle \mathbf{e_1} , \mathbf{e_2} \rangle = |\mathbf{e_1}|
|\mathbf{e_2}| \cos{\theta}.
\end{equation}
From (\ref{kappa_2}) it follows that the maximum probability is
reached at $|\cos{\theta}|=1$ and correspond to the value in
(\ref{kappa}).  Otherwise, in the case $ \langle \mathbf{e_1} ,
\mathbf{e_2} \rangle = 0$, there is always the trivial solution
$\mathbf{e_1}=\mathbf{e_2}=\mathbf{e_3}=\mathbf{e_4}=0$ and $E_5 =
\mathbb{I}$.


\begin{thebibliography}{99}

\bibitem{N-C} M. Nielsen, I. Chuang, \emph{Quantum Information and
Quantum Computation} Cambridge University Press, Cambridge (2000).

\bibitem{KLM} E. Knill, R. Laflamme, G. Milburn, \emph{A scheme for
efficient quantum computation with linear optics} Nature 409 46-52
(2001).

\bibitem{review} U. Leonhardt,
\emph{The physics of simple optical instruments} Rep. Prog. Phys. 66
1207-50 (2003).

\bibitem{review_1} P. Kok, W. J. Munro, K. Nemoto, T.~C.~Ralph,
J.~P.~Dowling and G.~J.~Milburn, \emph{Review article: Linear
optical quantum computation} quant-ph/0512071 (2005).

\bibitem{review_2} C. R. Myers, R. Laflamme,
\emph{Linear Optical Quantum Computation: an Overview}
quant-ph/0512104 (2005).

\bibitem{Cerf} N. J. Cerf, C. Adami, P. G. Kwiat, \emph{Optical
simulation of quantum logic} Phys. Rev. A 57 R1477-R1480 (1998).

\bibitem{P-R} P. Aniello, R. Coen Cagli, \emph{An Algebraic Approach
to Linear-Optical Schemes for Deterministic Quantum Computing} J.
Opt. B: Quantum Semiclass. Opt. 7 S711-S720 (2005).

\bibitem{future} P. Aniello, C. Lupo, M. Napolitano,
\emph{Exploring Representation Theory of Unitary Groups via Linear
Optical Passive Devices} to appear on Open Sys. Inf. Dyn.

\bibitem{JSmap} P. Jordan,
\emph{} Z. Phys. 94 531 (1935).

\bibitem{JSmap_1} J. Schwinger,
\emph{Quantum theory of angular momentum} L.~C.~Biedenharm and H.
Van Dam eds. (Academic Press) (1965).

\bibitem{Stone} M. H. Stone,
Proc. Nat. Acd. Scie. U.S.A. 16, 172-175 (1930).

\bibitem{vonN} J. von Neumann,
Math. Ann. 194, 570-578 (1931).

\bibitem{DiVi} D. DiVincenzo,
\emph{The Physical Implementation of Quantum Computation} Fortschr.
Phys. 48 771-783 (2000).

\bibitem{powers} S. Scheel, K. Nemoto, W.~J.~Munro and P.~L.~Knight,
\emph{Measurement-induced Nonlinearity in Linear Optics} Phys. Rev.
A 68, 032310 (2003).

\bibitem{eff-unit} G. G. Lapaire, P. Kok, J. P. Dowling and J.~E.~Sipe,
\emph{Conditional linear-optical measurement schemes generate
effective photon nonlinearities} Phys. Rev. A 68, 04234 (2003).

\bibitem{knill_2} E. Knill,
\emph{Quantum gates using linear optics and postselection} Phys.
Rev. A 66, 052306 (2002).

\bibitem{total} S. A. Gaal,
\emph{Linear Analysis and Representation Theory} Springer-Verlag,
Berlin (1973).

\bibitem{knill_1} E. Knill,
\emph{Bounds on the probability of success of postselected nonlinear
sign shitfs implemented with linear optics} Phys. Rev. A 68, 064303
(2003).

\bibitem{bplis}A.~Ferraro, S.~Olivares and M.~G.~A.~Paris,
  ``Gaussian States in Quantum Information'', {\em Napoli Series on
    Physics and Astrophysics} (Bibliopolis, Napoli, 2005).
\end{thebibliography}
\end{document}